\documentclass[pra,twocolumn,a4paper,showpacs,superscriptaddress]{revtex4-1}

 \usepackage{latexsym}
 \usepackage{amsmath}
 \usepackage{amsfonts}
 \usepackage{graphicx}
 \usepackage{amssymb}
 \usepackage{subfigure}
 \usepackage{color}

 \newcommand{\ket}[1]{\ensuremath{|#1\rangle}}
 \newcommand{\bra}[1]{\ensuremath{\langle #1 |}}

 \newcommand{\bc}{\begin{center}}
 \newcommand{\ec}{\end{center}}
 \newcommand{\mf}[1]{\boldsymbol{#1}}
 
 \newcommand{\ii}{i}
 \newcommand{\DP}{\Delta_1}
 \newcommand{\DC}{\Delta_2}
 \newcommand{\WP}{\omega_1}
 \newcommand{\WC}{\omega_2}

\begin{document}

\title{Optical cloning of arbitrary images beyond the diffraction limits}

\author{Onkar N. \surname{Verma}}
\email{onkar@iitg.ernet.in}
\affiliation{Department of Physics, Indian Institute of Technology Guwahati, Guwahati- 781 039, Assam, India}

\author{Lida \surname{Zhang}}
\email{lida.zhang@mpi-hd.mpg.de}

\author{J\"org \surname{Evers}}
\email{joerg.evers@mpi-hd.mpg.de}
\affiliation{Max-Planck-Institut f\"ur Kernphysik, Saupfercheckweg 1, 69117 Heidelberg, Germany}

\author{Tarak N. \surname{Dey}}
\email{tarak.dey@gmail.com}
\affiliation{Department of Physics, Indian Institute of Technology Guwahati, Guwahati- 781 039, Assam, India}
\affiliation{Max-Planck-Institut f\"ur Kernphysik, Saupfercheckweg 1, 69117 Heidelberg, Germany}
\date{\today}

\begin{abstract}
Cloning of arbitrary images encoded onto the spatial profile of a laser beam onto that of  a second beam is theoretically investigated. The two fields couple to an atomic lambda system in a coherent population trapping configuration. In particular, the case in which the probe and control fields are of comparable strength is considered. By considering more and more complex structures, we eventually find that our method is suitable to clone arbitrary images, which we demonstrated by a full numerical simulation of the propagation dynamics  of both applied fields in the atomic medium, with the three letters ``CPT" encoded on the initial control field profile. We find that the cloned structures have feature sizes reduced by about a factor of 2 compared to the initial images, consistent with a recent related experiment.
\end{abstract}
\maketitle
\section{Introduction}
Optical diffraction forms a fundamental limitation to the creation and detection of small images for conventional optical devices and microscopy. The reason is that any image with finite size can be considered as a group of different plane-wave components. Each component acquires a different phase shift during propagation either in free space or in medium. The resulting superposition of all wave components leads to diffraction, which means that the transmitted image will be distorted even after  propagating only a few Rayleigh lengths~\cite{Rayleigh}. From this interpretation it is clear that diffraction can be eliminated if conditions are tailored such that each plane-wave component acquires the same phase shift during propagation. To achieve this, it has for example been suggested that several specially-shaped beams with spatially-dependent phases such as Airy~\cite{berry1979,siviloglou2007}, Bessel~\cite{durnin1987,mcgloin2005}, and Mathieu beams~\cite{vega2000,zhang2012}, can propagate without 
diffraction in free space, since the phase differences between the different components are exactly 
compensated by the initially spatially dependent phases.    

It has also been recognized that optical diffraction can be greatly suppressed or even eliminated by using atomic coherence effects such as electromagnetically induced transparency \cite{Moseley1, Moseley2,Truscott,Kapoor,Yavuz,Gorshkov,Fleischhauer}, coherent population trapping (CPT)~\cite{Vudyasetu,Kapale,Kiffner,Dey1}, or saturated absorption techniques~\cite{Dey2} in multilevel atomic systems. Most of these methods can be understood by noting that the spatially-dependent probe and coherent control fields couple to the atomic transition in such a way, that the probe field experiences a spatially dependent refractive index depending on the intensity of the control fields. As an example, a suitable spatially dependent control field can optically induce a waveguide-like structure. When the probe field propagates in the induced waveguide inside the atomic medium, it is mostly confined in the waveguide due to total reflection, resulting in the elimination of the diffraction for the probe. 

A more interesting and counterintuitive approach, which is fundamentally different from previous methods, has been developed by Firstenberg {\it et al.} \cite{Firstenberg1,Firstenberg2}. They have theoretically proposed and experimentally verified that atomic motion and collisions can be utilized to eliminate the paraxial diffraction of arbitrary image. The underlying physics is that the phase difference acquired during propagation for each plane-wave component of the image is exactly compensated by an additional phase shift induced by the atomic motion, thus leading to the elimination of diffraction. However, this method is plagued by strong absorption.

Recently, Li {\it et al.} \cite{Li} experimentally demonstrated that the spatial shape of a control beam can be cast onto a weak probe beam via coherent population trapping in a three-level lambda atomic system. In their experiment, the transmitted intensity of the probe beam had a similar spatial profile as that of the control beam, no matter what the input probe is. Moreover, the size of the transmitted probe beam was half of that of the diffraction-limited input probe. Manipulation of the susceptibility along the transverse direction is the key idea behind mapping the spatial shape of the control beam onto the probe beam. It has been used for focusing, defocusing, and self-imaging \cite{Moseley1, Moseley2,Cheng,lida}.

It is important to note that the probe field is typically treated as a weak field and the control field as a strong field, such that perturbation theory can be employed to derive the linear effect of the atomic medium on the probe field propagation. There have been theoretical and experimental studies where the probe is not necessarily weak \cite{Greentree,Strekalov,Tewari}.
In this situation, the effect of the atomic coherence on the propagation dynamics of the control field need to be taken into account \cite{Dey3}.

In this paper, we theoretically investigate the possibility of cloning of an arbitrary image carried on an control field to another probe beam. The two control and probe beams are coupled to a three-level atomic lambda system to form a CPT configuration. We assume the two laser fields to be of comparable strength, such that perturbation theory for the probe field  is not valid any more to describe the effect of the atomic medium on the two fields. 
We start by calculating the susceptibilities including linear and nonlinear effects for both fields by solving the related density-matrix equations. 
As expected, we find that a spatially-dependent refractive index for the probe field is generated, structured by the spatial intensity profile of the control beam. In particular, the generated structures enable one to transfer the transverse distribution of the control field onto the transmission profile of the probe field. 
In order to study the full propagation dynamics, we then numerically solve the paraxial propagation equations for both fields by using a higher order split operator method~\cite{Shen}. 
We begin our analysis with a Gaussian control and a super-Gaussian probe field and observe the gradual mapping of the control field onto the probe field throughout the propagation. We in particular find that in the case of a strong probe field, the transmitted probe beam is focused more tightly by a factor of two compared to the weak probe field case. 
Next, we consider a control field with a spatial two-peaked Hermite-Gaussian profile, and demonstrated cloning of the profile onto the probe beam with feature size reduced by a factor of about 2.5. 
In order to verify that our method can serve as an universal tool for cloning of arbitrary image, we finally simulate the three-dimensional light propagations for both fields, in which the spatial profile of the control field carries the three letters ``CPT". We show that also this structure can be cloned onto the probe beam which initially has a simple plane-wave profile, even though the control field is severely distorted throughout the propagation due to diffraction. Again, we observe a reduction of the feature size by a factor of about 2 in the probe field.

The article is organized as follows. In Sec. II A, we introduce our theoretical model and then employ a semi-classic theory to describe the system. In Sec. II B, we discuss the propagation equations for all fields in the strong field limit. In Sec. II C, we analytically calculate the atomic responses to the probe and control fields, including both linear and high-order nonlinear effects. In Sec. II D, we specify the spatially-dependent profiles for both fields that are used in our numerical simulation. In Sec. III A, we first show the differences in the spatial dependence of the absorption properties for the probe field in the weak and strong field limit, respectively. Then, we discuss the spatial refractive modulation which is dependent on the probe detuning. Finally, the refractive index for the control is investigated in both weak and strong field cases. In Sec. III B, we discuss our numerical results on the  light propagation in several cases where different spatial profiles are chosen for both fields.  
Sec. IV provides a brief summary of our paper.

\section{Model}
\subsection{Equations of motion}
\begin{figure}[t]
\includegraphics[width=0.8\columnwidth]{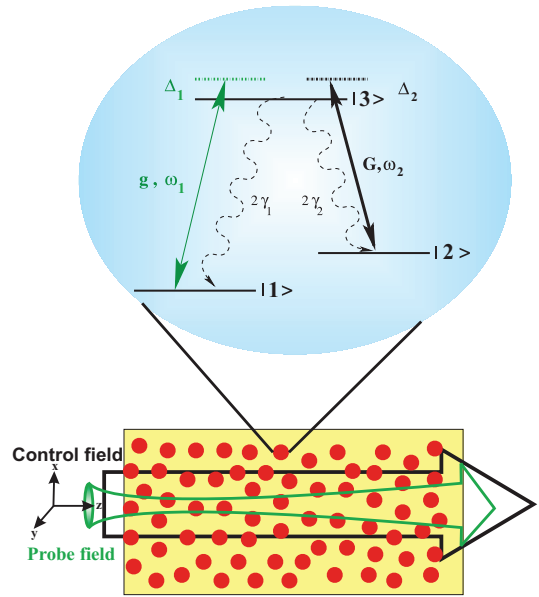}
\caption{\label{fig:Fig1} (Color online) Schematic setup for the cloning of arbitrary images encoded in the spatial profile of a control field onto the spatial profile of a probe beam. Both fields co-propagate through a rubidium vapor cell, and couple to the atoms on the $|3\rangle\leftrightarrow|1\rangle$ (probe field with frequency $\omega_1$)  and $|3\rangle\leftrightarrow|2\rangle$ (control field with frequency $\omega_2$) transitions, respectively. Our analysis includes the case in which both fields are strong.}
\end{figure}

We consider a homogeneous cloud of $^{87}$Rb atoms, with two laser fields coupled to the atoms such that a three-level $\Lambda$ scheme as shown in Fig.~\ref{fig:Fig1} is of relevance. The probe field couples to transition $|1\rangle \leftrightarrow |3\rangle$, and the control field to transition $|2\rangle \leftrightarrow |3\rangle$. The two co-propagating fields are defined as
\begin{equation}
\label{field}
 {\vec{E}_j}(\vec{r},t)= \hat{e}_{j}\mathcal{E}_{j}(\vec{r})~e^{- i\left(\omega_j t-  k_j z\right )} + {c.c.}\,,
\end{equation}
where $\mathcal{E}_{j}(\vec{r})$ are the slowing varying envelope functions, $\hat{e}_{j}$ the unit polarization vectors, $\omega_i$ the laser field frequencies and $k_j$ is the wave numbers of fields.  The index $j\in \{1,2\}$ denotes the probe or control field, respectively. 

The Hamiltonian of the system in electric dipole and rotating wave approximation is given by,
\begin{align}
 H  = & \hbar\omega_{31}\ket{3}\bra{3} + \hbar\omega_{32}\ket{2}\bra{2}  \notag \\
 & -\hbar \left(ge^{-i\WP t}\ket{1}\bra{3} + Ge^{-i\WC t} \,\ket{2}\bra{3}    \,+\,\text{H.c.}\right)\,,
 \label{H}
\end{align}
with Rabi frequencies of the probe and control fields defined as
\begin{subequations}
\label{Rabi_fre}
\begin{align}
g&=\frac{\vec{d}_{31}\cdot\vec{\mathcal{E}}_{\rm{1}}e^{ik_1 z}}{\hbar}\,,\\
G&=\frac{\vec{d}_{32}\cdot\vec{\mathcal{E}}_{\rm{2}}e^{ik_2 z}}{\hbar}\,.
\end{align}
\end{subequations}
The atomic transition frequencies and the corresponding dipole moment matrix elements are denoted by $\omega_{3j}$ and $\mf{\hat{d}}_{3j}$, respectively, and we label the radiative decay rate from state $|3\rangle$ to ground state $|j\rangle$ by $2\gamma_{j}$.  The master equation for the density operator $\rho$ is given by 
\begin{align}
\label{master}
\dot{\rho}=-\frac{\ii}{\hbar}\left[H,\rho\right]+\mathcal{L}_{\gamma}[\rho].
\end{align}
The last term in Eq.~(\ref{master}) describes incoherent processes such as spontaneous emission and is determined by
\begin{align}
\mathcal{L}_{\gamma}[\rho] = &-\gamma_{1}\left(\ket{3}\bra{3}\rho-2\ket{1}\bra{1}\rho_{33}+\rho\ket{3}\bra{3}\right)  \notag \\
 & -\gamma_{2}\left(\ket{3}\bra{3}\rho-2\ket{2}\bra{2}\rho_{33}+\rho\ket{3}\bra{3}\right)\,.
 \label{decay}
\end{align}
In a suitable interaction picture, the density matrix equations follow as 
\begin{subequations}
\label{Full_density}
\begin{align}
 \dot{\sigma}_{33}=&-2(\gamma_{1}+\gamma_{2})\sigma_{33}+ \ii G \sigma_{23}
 + \ii g \sigma_{13} \nonumber \\
& -\ii G^{*} \sigma_{32}-\ii g^{*} \sigma_{31} \,,\\
 \dot{\sigma}_{22}=& 2 \gamma_{2}\sigma_{33} + \ii G^* \sigma_{32}  - \ii G
 \sigma_{23}\,,\\
 \dot{\sigma}_{32}=&-\left[\frac{1}{2}(\gamma_{1} + \gamma_{2}) + \ii \DC\right]\sigma_{32} + \ii g \sigma_{12} \nonumber\\
&+ \ii G  \sigma_{22} - \ii G\sigma_{33} \,,\\
 \dot{\sigma}_{31}=&-\left[\frac{1}{2}(\gamma_{1} + \gamma_{2}) + \ii \DP\right]\sigma_{31}
 + \ii G\sigma_{21} \nonumber\\
& + \ii g\sigma_{11} - \ii g\sigma_{33}\,,\\
 \dot{\sigma}_{21}=& -[\Gamma- \ii (\DC-\DP)]\sigma_{21}
 + \ii G^*\sigma_{31} - \ii g\sigma_{23}\,.
 \end{align}
\end{subequations}
The detunings of the probe and the control fields from the respective transition frequencies are defined as
\begin{subequations}
\begin{align}
 \DP &= \omega_{31} - \omega_1\,,\\
  \DC &= \omega_{32}- \omega_2\,.
\end{align}
\end{subequations}
We have further included pure dephasing of the ground state coherence, e.g., due to phase changing collisions, and denote the total decay rate of the coherence by $\Gamma$.
The remaining density matrix equations follow from the constraints $\sigma_{11}+\sigma_{22}+\sigma_{33}=1$ and $\sigma_{ij}=\sigma_{ji}^*$.

\subsection{Propagation equation}
We use Maxwell's wave equations to simulate the spatial evolution of the control and the probe beams through the medium, in order to study the effect of both diffraction and dispersion during the propagation. The wave equations for the probe ($j=1$) and control ($j=2$) fields can be written as
\begin{align}
\label{Maxwell}
 \left(\vec{\nabla}^2-\frac{1}{c^2}\frac{\partial^2}{\partial t^2}\right)\vec{E}_j=\frac{4\pi}{c^2}\frac{\partial^2\vec{\mathcal{P}}_j}{\partial t^2},
\end{align}
where $\vec{\mathcal{P}}_j$ are the macroscopic polarizations induced by the control and probe fields, respectively. They can be expressed in terms of both the atomic coherences as well as the susceptibility as
\begin{align}
\label{polarization}
\vec{\mathcal{P}}_j&=\mathcal{N}\left(\vec{d}_{3j}\sigma_{3j}e^{-\ii\omega_j t}+c.c.\right)\notag\\
&=\left(\chi_{3j} \hat{e}_{j}\mathcal{E}_je^{-\ii\omega_j t}+c.c.\right)\,,
\end{align} 
where $\mathcal{N}$ is the density of the atomic medium, and $\chi_{31}$ and $\chi_{32}$  are the susceptibilities for the response to the probe and control fields, respectively. 
In slowly varying envelope and paraxial wave approximation,  Eqs. (\ref{field}), (\ref{Rabi_fre}) and (\ref{Maxwell}) lead to propagation equations for the two fields given by 
\begin{subequations}
\label{probe}
\begin{align}
 \frac{\partial g}{\partial z}
   &= \frac{\ii }{2{k_1}} \left( \frac{\partial^2 }{\partial x^2}
      + \frac{\partial^2 }{\partial y^2} \right) g + 2i{\pi}k_1{\chi}_{31}\,{g} \,, \\
 \frac{\partial G}{\partial z}
   &= \frac{i}{2{k_2}} \left( \frac{\partial^2 }{\partial x^2}
      + \frac{\partial^2 }{\partial y^2} \right) G +2i{\pi}k_2{\chi}_{32}\,{G}\,,
\end{align}
\end{subequations}
The first terms in the parentheses on the right hand sides account for the diffraction. The second terms on the right hand sides are responsible for the dispersion and absorption of the both the control and probe beams. Note that the two propagation equations are coupled via the susceptibilities $\chi_{31}$ and $\chi_{32}$.

\subsection{Medium susceptibility}
Next we calculate the response of the medium to the probe and control fields, characterized by the respective susceptibilities. In steady state, the atomic coherences $\sigma_{31}(\omega_p)$ and $\sigma_{32}(\omega_c)$ are obtained from Eqs.~(\ref{Full_density}) as
\begin{equation}
\label{atomic coherence}
 {\sigma}_{3j}=\frac{N_{3j}}{D}\,,
\end{equation} 
where the numerators $N_{32}$, $N_{31}$ and the denominator $D$ are listed in Appendix~\ref{app-A}. The expressions are rather complex, since we include the fields to all orders, in order to account for nonlinear effects. To simplify the expressions, we have assumed equal decay rates on the two transitions, $\gamma_{1}=\gamma_{2}=\gamma$.
Using Eqs.~(\ref{polarization}), one can readily obtain the susceptibility at the frequencies at $\omega_c$ and $\omega_p$ as
\begin{subequations}
\label{chis}
\begin{align}
 \label{chi_32}
 {\chi}_{32}(\omega_c)&=\frac{\mathcal{N}|d_{32}|^2}{{\hbar}G}{\sigma}_{32}\,, \\
  \label{chi_31}
  {\chi}_{31}(\omega_p)&=\frac{\mathcal{N}|d_{31}|^2}{{\hbar}g}{\sigma}_{31} .
\end{align}
\end{subequations}

\subsection{Beam profiles}
In the main part of our result section, we will numerically propagate complex transverse beam profiles. But first, in order to interpret the effect of the beam profiles on the propagation, we chose the transverse spatial profile of the control field as a Hermite-Gaussian. At $z=0$, it can be written as,
\begin{align}
\label{shape_con}
G(x,y)=&{G}_0\:H_m \left( \frac{x\sqrt{2}}{w_c} \right)H_n \left( \frac{y\sqrt{2}}{w_c} \right)
e^{-\frac{(x^2+y^2)}{w_c^2}}\,.
\end{align}
Here, $G_0$ is the input amplitude, and $w_c$ is the width of the control field. The function $ H_k $ is a Hermite polynomial of order $k$,  and the indices $m$ and $n$ determine the shape of the control field profile along the $x$ and $y$ directions, respectively. Since we want to consider the transfer of arbitrary spatial information, we will study different values of $m,n$ in the following.

Similarly, the probe field is initially assumed to have a super-Gaussian transverse profile given by
\begin{align}
\label{shape_probe}
g(x,y)=&{g}_0 \:e^{-\frac{ (x^2+y^2)^8}{w_p^{16}}}\,.
\end{align}
The initial peak amplitude and the width of the probe field are denoted by $g_0$ and $w_p$,  respectively.
Instead of choosing super Gaussian as an initial profile, the shape of the probe field can be consider any arbitrary shape such as a plane wave, Gaussian or hyperbolic shape. 
The desired spatial profile of the probe beam can be generated by using a spatial light modulator based on either liquid crystal or coherent EIT media\cite{Bourderionnet,Zhao}. 

\section{Results}

\subsection{Linear and nonlinear susceptibility }

\begin{figure}[b]
\includegraphics[width=0.8\columnwidth]{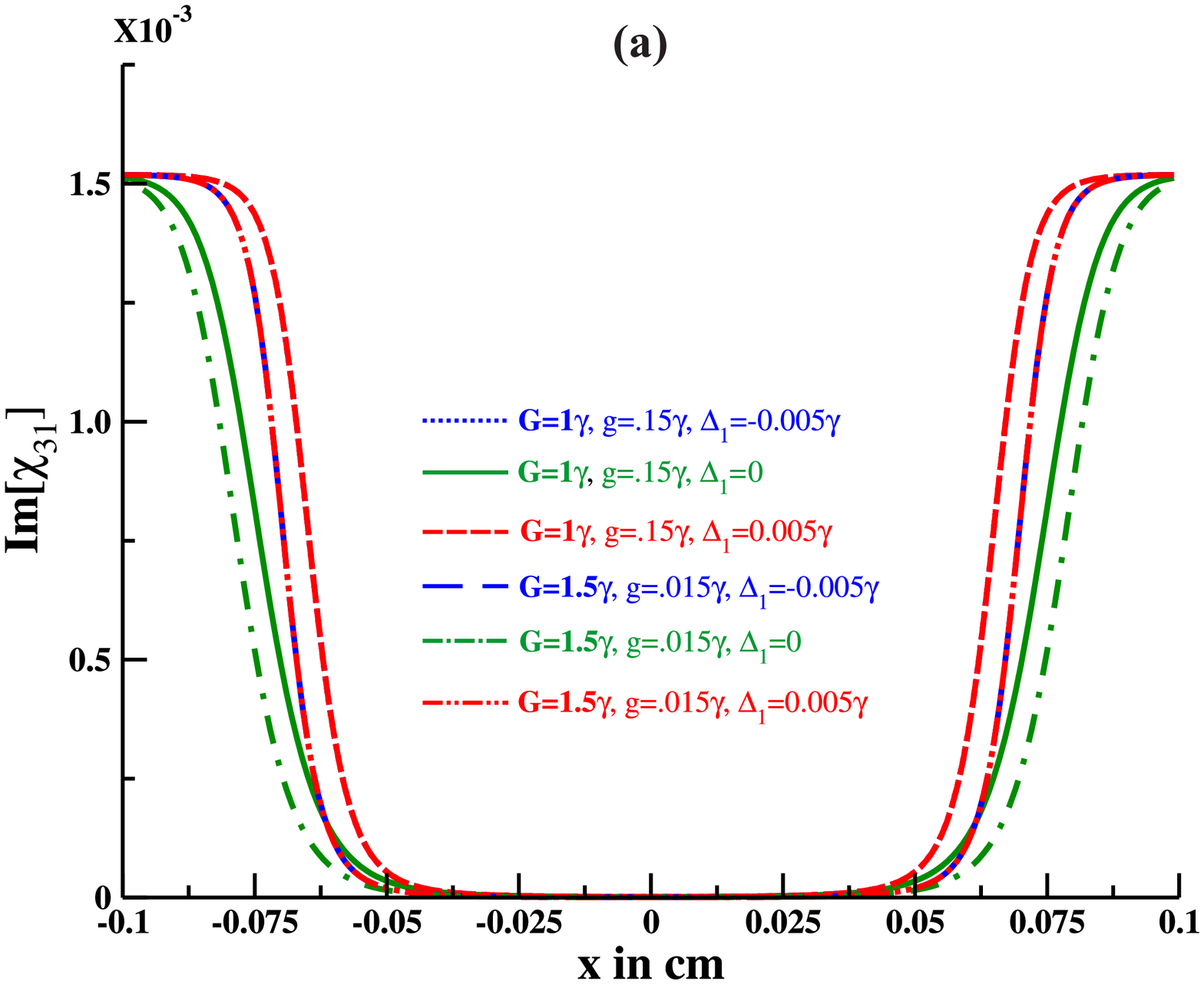}
\includegraphics[width=0.8\columnwidth]{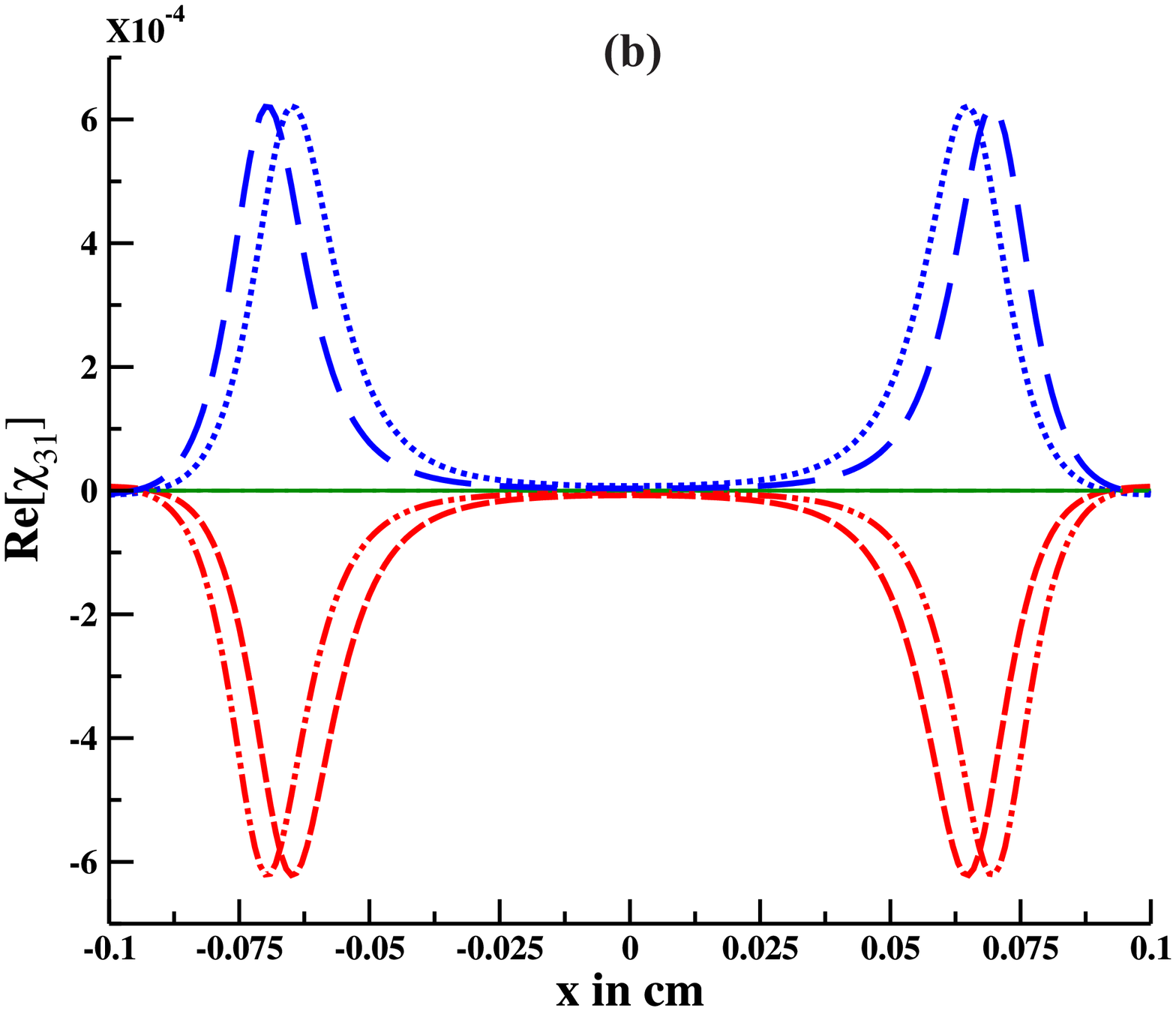}
\caption{\label{fig:Fig2} (Color online) Probe beam susceptibility as a function of the transverse coordinate $x$. The control field has a Gaussian $(m=0,n=0)$ beam profile. (a) shows the imaginary and (b) the real component of the susceptibility. The three different curves show probe field detunings  $\DP=-0.005$ (blue dotted line, blue dashed line), $\DP=0.0$ (green solid line, green dot-dashed line), and $\DP=0.005{\gamma}$ (red dashed line, red dashed double-dot line) for two data sets of probe amplitudes $g=0.015{\gamma}$ and $g=0.15{\gamma}$, respectively.
The other parameters are $y=0.005$cm, ${w_p}=150{\mu}$m, ${w_c}=400{\mu}$m, $\DC=0$, $\Gamma=0.001\gamma$, and $N=5{\times}10^{11}$ atoms/cm$^{3}$.}
\end{figure}

We now turn to our results, and start by studying the probe and control field susceptibilities at frequencies $\WP$ and $\WC$ using Eqs.~(\ref{chis}). 
For this purpose, we first consider a super Gaussian probe field and a Gaussian control field [$m=0,n=0$ in Eq.~(\ref{shape_con})] such that the probe susceptibility becomes inhomogeneous along the transverse directions. 
Results are presented for two different cases of the initial field amplitudes. We denote the case $g_0 \ll G_0$ as the weak field limit, and $g_0 \sim G_0$ as the strong field case. In our numerical calculations, we choose $g_0 = 0.015\,\gamma$ and $g_0=0.15\,\gamma$ for the two cases, respectively.

Fig.~\ref{fig:Fig2} displays the spatial dependence of the probe field susceptibility. The different curves show Im$(\chi_{13})$ and Re$(\chi_{13})$  for three different detunings and both probe intensity cases as a function of $x$, with $y=0.005$~cm. From panel (a), we find that the probe field is essentially transmitted without absorption in regions in which the control field amplitude is large, whereas in the wings ($ |x| > $ 0.05 cm) with low control field, strong absorption occurs. 
As expected, we found that the width of the transparency window depends on the intensities of the control and probe fields as well as the decay rate of ground state atomic coherence. 
Comparing the two intensity cases, we find that a relative increase of the probe field intensity results in steeper transitions between regions of high and low absorption. Later, we will show that this in turn leads to a smaller feature size  of the image cloning scheme. 

The transparency of the medium to the probe field can be understood as arising from coherent population trapping (CPT) or electromagnetically induced transparency (EIT), depending on the relative strength of the two applied fields. In both cases, destructive interference occurs between two different excitation pathways from $|1\rangle$ or $|2\rangle$ to the excited state $|3\rangle$. CPT and EIT are restricted to a certain transparency window. This transparency window can be controlled via the external parameters, and permits to transfer the transverse distribution of the control field onto the transmission profile of the probe field. This is the key mechanism of cloning the control field profile to the probe field.

The corresponding real part of the susceptibility is shown in Fig.~\ref{fig:Fig2}(b). We notice that it can be controlled between negative and positive values via the detuning. For red detuning of the probe field ($\DP=0.005{\gamma}$), it has a local maximum around $x=0$, whereas for blue detuning, a minimum occurs around $x=0$. For non-zero detunings, the real part becomes spatially dependent in particular around the regions in which the absorption undergoes a transition from low to high absorption. 

Therefore, at red detuning, the  transverse profile of the control field allows one to imprint a fiber-like refractive index gradient onto the atomic medium. This parabolic refractive index  variation causes focusing of the probe field towards the center of the control field and also guides the probe field propagation along the propagation axis. 
Conversely, at  blue detuning of the probe field, the real part  has maxima at its wings, which is referred to as anti-waveguide like refractive index \cite{Bortman-Arbiv}.  This anti-waveguide refractive index leads to shape distortions of the probe field. Hence, electromagnetically induced focusing and defocusing is possible in our setup by properly detuning the probe field. 
At two photon resonance condition, {\it i.e.}, $\DP=\DC=0$, only the imaginary part of the susceptibility varies spatially, while the real part is constant. Then the probe field propagates through the transparency window without being focused or defocused. Note that the probe field will nevertheless spread due to the inevitable diffraction. 

\begin{figure}[b]
\includegraphics[width=0.8\columnwidth]{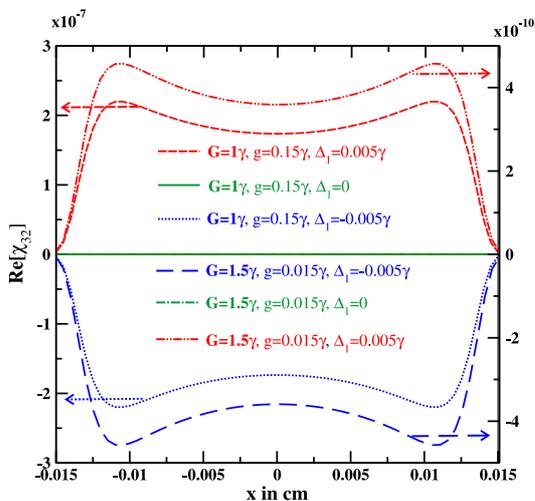}
\caption{\label{fig:Fig3} (Color online) Real part of the control field susceptibility as a function of transverse position $x$. The control field has a Gaussian $(m=0,n=0)$ shape. The three different curves show probe field detunings  $\DP=-0.005$ (blue dotted line, blue dashed line), $\DP=0.0$ (green solid line, green dot-dashed line), and $\DP=0.005{\gamma}$ (red dashed line, red dashed double-dot line) for two data sets of probe amplitudes $g=0.015{\gamma}$ and $g=0.15{\gamma}$, respectively, with different scales of the axes.
The other parameters are same as in Fig(\ref{fig:Fig2}).}
\end{figure}

Next, we consider the effect on the control field. Figure~\ref{fig:Fig3} shows the real part of the susceptibility experienced by the control field. It in particular illustrates that it becomes especially important when the relative intensity of the probe and control fields are of comparable strength, as for the intense probe field case, the spatial dispersion of the control field is about three orders of magnitude larger than in the weak probe field case. Hence, it is important to include atomic coherence effects in the paraxial wave equation for the control field. It is also evident from Fig.~\ref{fig:Fig3} that in addition to the diffraction, the  variation of the refractive index  induces further distortions of the control field shape for both red- and blue-detuned probe fields.   
Our further analysis also showed that the absorption of the control field is rather low as compared to the probe field absorption in both cases, such that the control field can be propagated through the medium without significant absorption for relevant parameter ranges. 

\begin{figure}[t]
\includegraphics[width=0.8\columnwidth]{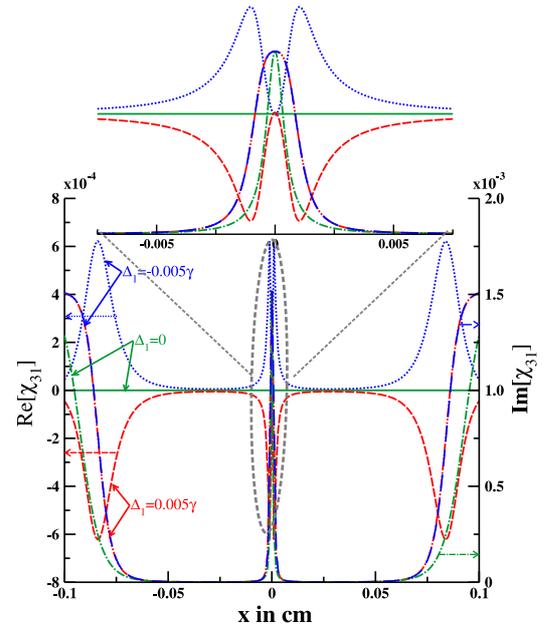}
\caption{\label{fig:Fig4} (Color online) Probe field susceptibility as a function of the transverse spatial coordinate $x$. 
The control field has a doubly-peaked spatial profile characterized by $m=1$ and $n=0$ in Eq.~(\ref{shape_con}).
The central part around $x=0$ is magnified in the inset in the top part of the figure.
The other parameters are $y = 0.005$ cm, $G_0=1.5\gamma$, $g_0=0.015\gamma$, $w_p=400{\mu}$m and $N=7.5{\times}10^{11}$ atoms/cm$^{3}$.}

\end{figure}
\begin{figure}[t]
\includegraphics[width=0.8\columnwidth]{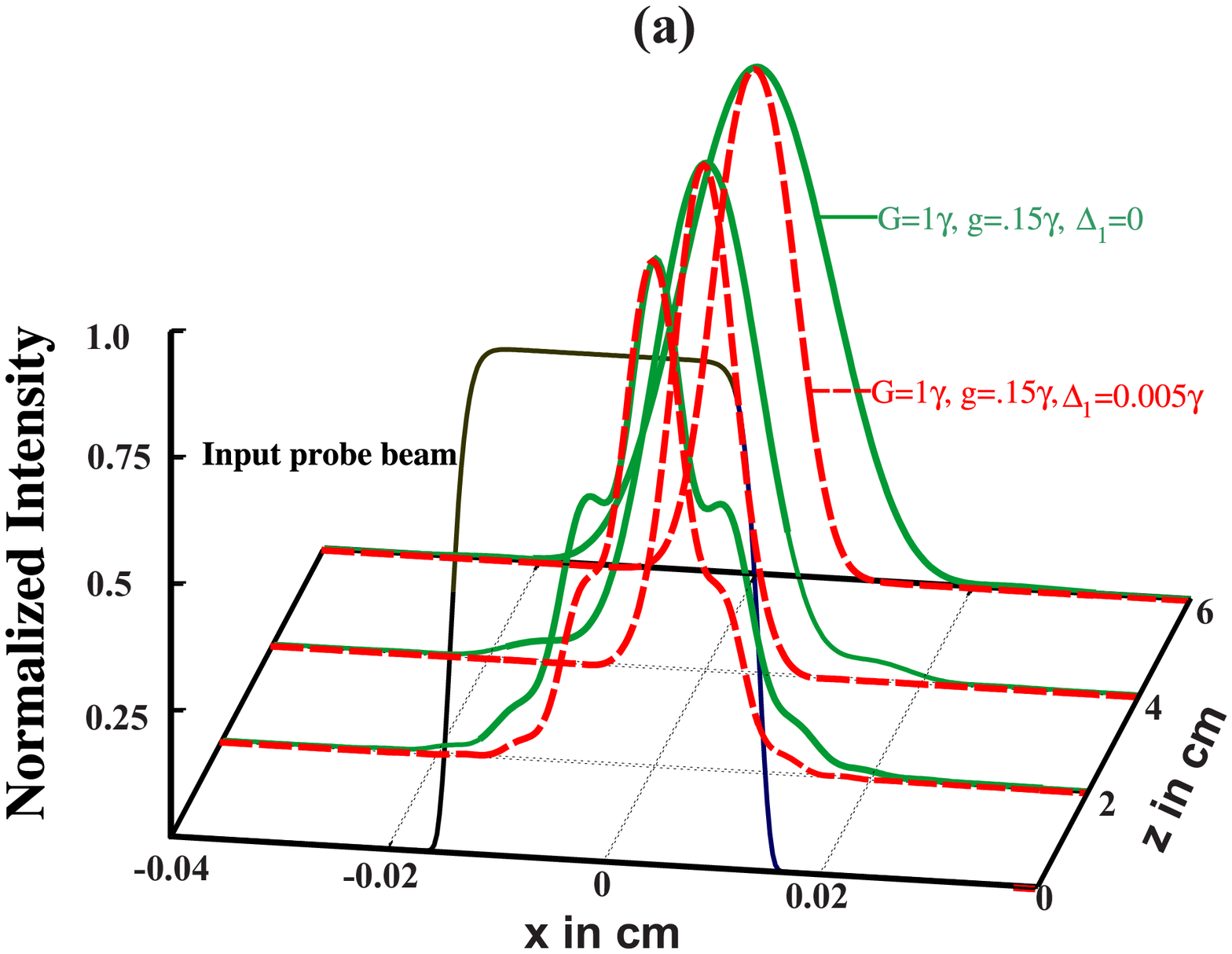}
\includegraphics[width=0.8\columnwidth]{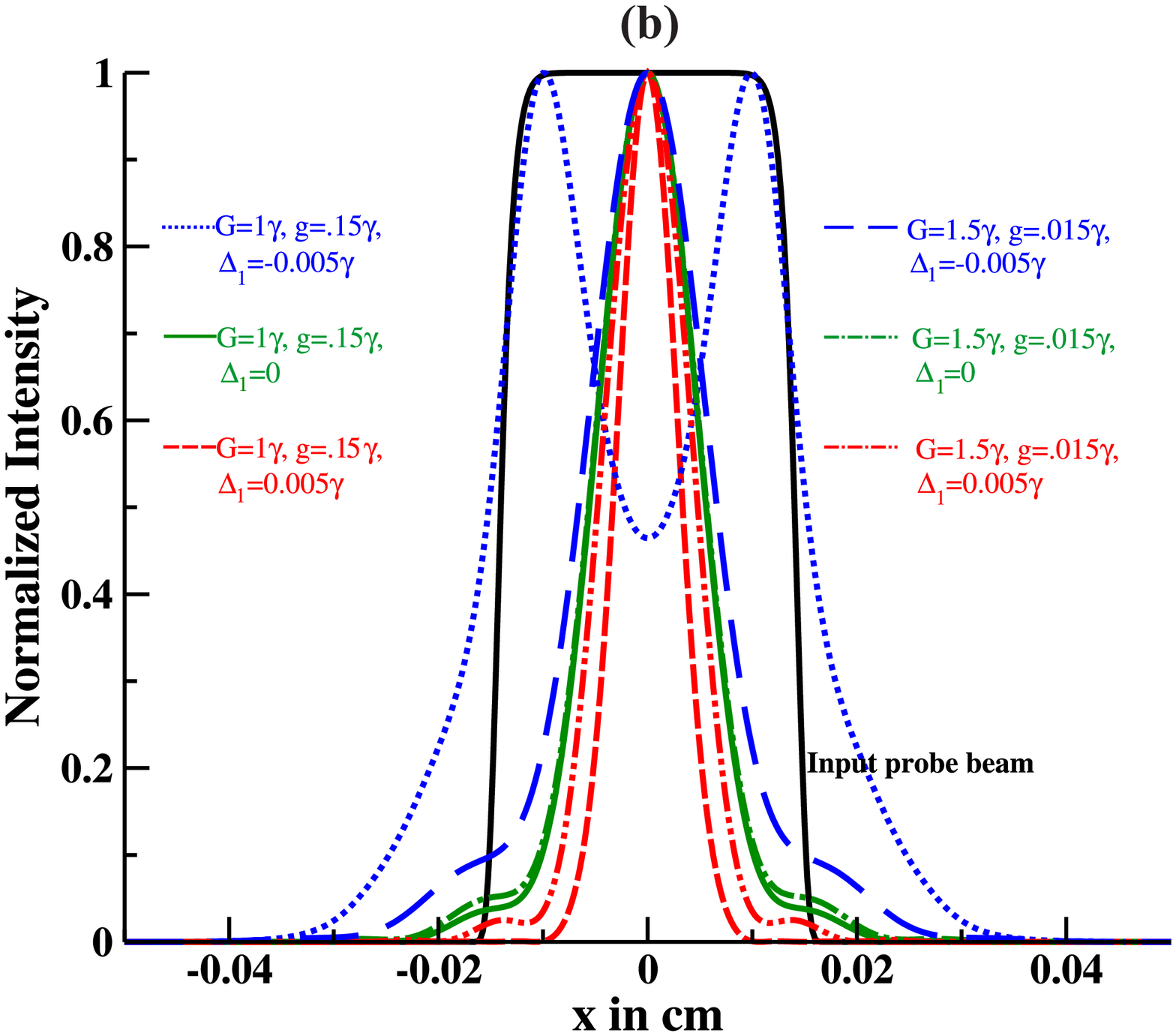}
\caption{\label{fig:Fig5} (Color online) Normalized intensity profile of the propagating probe beam against the transverse coordinate $x$ for $y=0$. (a) shows the beam profile at different propagation distances $z$. In (b), the transmitted probe beam intensity at the output of a 4~cm long medium is shown for different detunings. 
All other parameters are chosen as in Fig.~\ref{fig:Fig2}.}
\end{figure}
\begin{figure}[b]
\includegraphics[width=0.8\columnwidth]{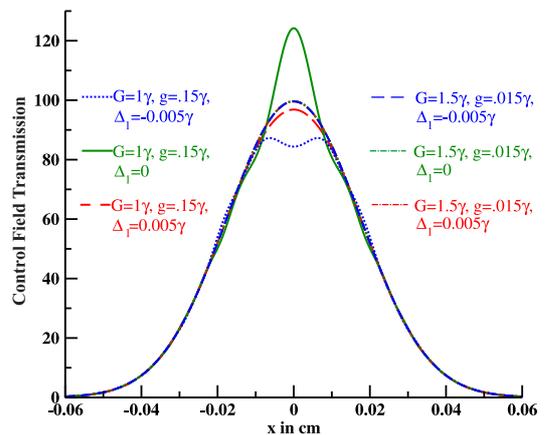}
\caption{\label{fig:Fig6} (Color online) Spatial intensity profile of the control field as a function of the transverse coordinate $x$ after propagation through a 4~cm long atomic medium. 
Results are shown for different field parameters. 
The profile is shown in the $y=0$ plane. All other parameters are the same as in Fig.~\ref{fig:Fig5}.}
\end{figure}

Finally, we analyze the dependence of the probe susceptibility on the control field transverse beam shape. Figure~\ref{fig:Fig4} shows the transverse variation of the probe field susceptibility for a doubly peaked transverse spatial distribution of the control field obtained by setting $m=1$ and $n=0$ in Eq.~(\ref{shape_con}). The central part of the figure is shown enlarged in the inset. Overall, the interpretation is similar to the case of a singly-peaked control field beam profile. But the doubly-peaked structure leads to the formation of two transparency windows. For positive probe detuning ($\DP=0.005$), each of the transparency windows features a fiber-like refractive index, focusing the probe field towards the waist of the control field and thereby reducing the feature size. The generalization of these results to other spatial modes of the control field with different values of $m,n$ is straightforward.

\subsection{Propagation dynamics of probe and control beams}

In this Section, we present results for the evolution of the spatial beam profile throughout the propagation. For this, we numerically integrate the full set of paraxial wave equations Eqs. (\ref{probe}) using a higher order split operator method~\cite{Shen}. 
We begin with a super-Gaussian probe beam shape and a Gaussian control beam shape to demonstrate the mapping of the control field profile onto the probe field. Results are shown in Fig.~\ref{fig:Fig5}. Panel (a) depicts the intensity profile of the probe beam at different propagation distances $z$. The initial field amplitudes are $G_0=1\gamma$ and $g_0=0.15\gamma$, corresponding to the strong probe field case. We find that the probe beam profile gradually acquires the control beam shape as it propagates through the medium. This process is completed at approximately $z=2$~cm. Interestingly, the probe field width at this propagation distance is about $60{\mu}$m which is an order of magnitude less than the initial width of the  control beam. 
At this distance, the Rayleigh length of the cloned probe beam is about 1.42 cm. Consequently, the cloned beam suffers from diffraction throughout the further propagation. We also notice from Fig.~\ref{fig:Fig5}  that at two photon resonance condition $\DP = \DC = 0$, the probe beam experiences diffraction. In contrast, at red detuning, the probe field propagates unaltered through the medium. We thus find that  the probe field diffraction can be controlled by a proper choice of the single photon detuning $\DP$ of the probe field. 

Fig.~\ref{fig:Fig5}(b) further illustrates how the intensity and detuning of the probe field can be used to control the width of the transmitted probe beam. It shows that the probe beam is more tightly focused by a factor of about two for the strong control field case compared to the weak probe field case. At red detuning condition, our numerical computations show that the width and the transmission of the output probe beam are $85{\mu}$m [$55{\mu}$m] and $20\%$ [$5\%$] for weak [strong] field case. 

Figure~\ref{fig:Fig6} show the corresponding results for the spatial variation of the transmitted control beam intensity profile in the $y=0$ plane after propagation through a 4 cm long atomic medium. For the weak probe field case, the control field propagates essentially without any distortion. But in the strong probe field case, we find from Fig.~\ref{fig:Fig6} that the control field shape is distorted. Nevertheless, the integrated transmission intensity of  the control field is approximately $100\%$ in both cases. Thus, we conclude that the control beam intensity is spatially redistributed in the strong probe field case. As a consequence, it is necessary to include the effect of atomic coherence in the propagation of the control field in the non-resonant as well as in the resonant condition for the strong probe field case.
\begin{figure}[t]
\centering
\includegraphics[width=0.6\columnwidth,angle=-90]{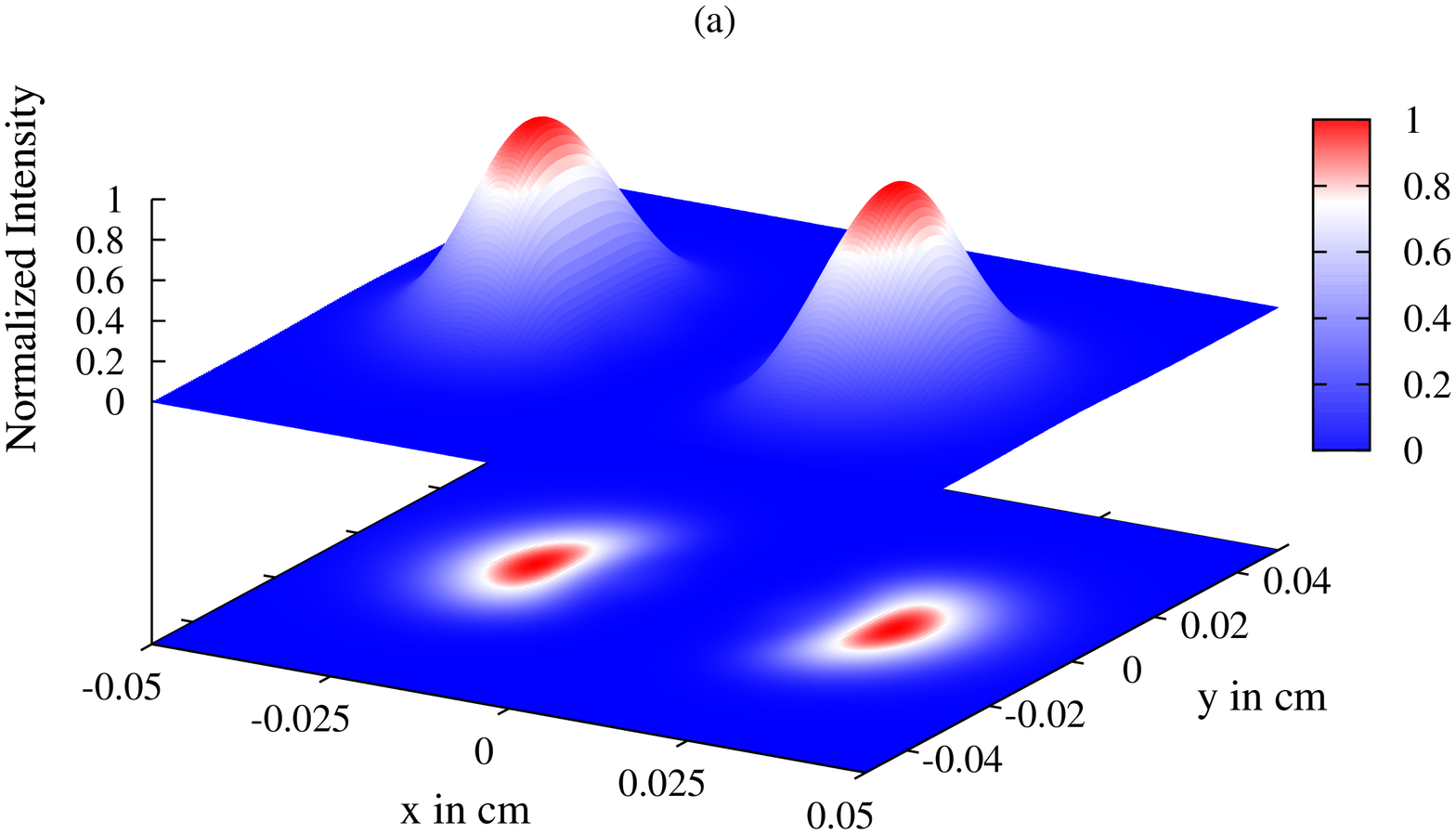}\\
\includegraphics[width=0.6\columnwidth,angle=-90]{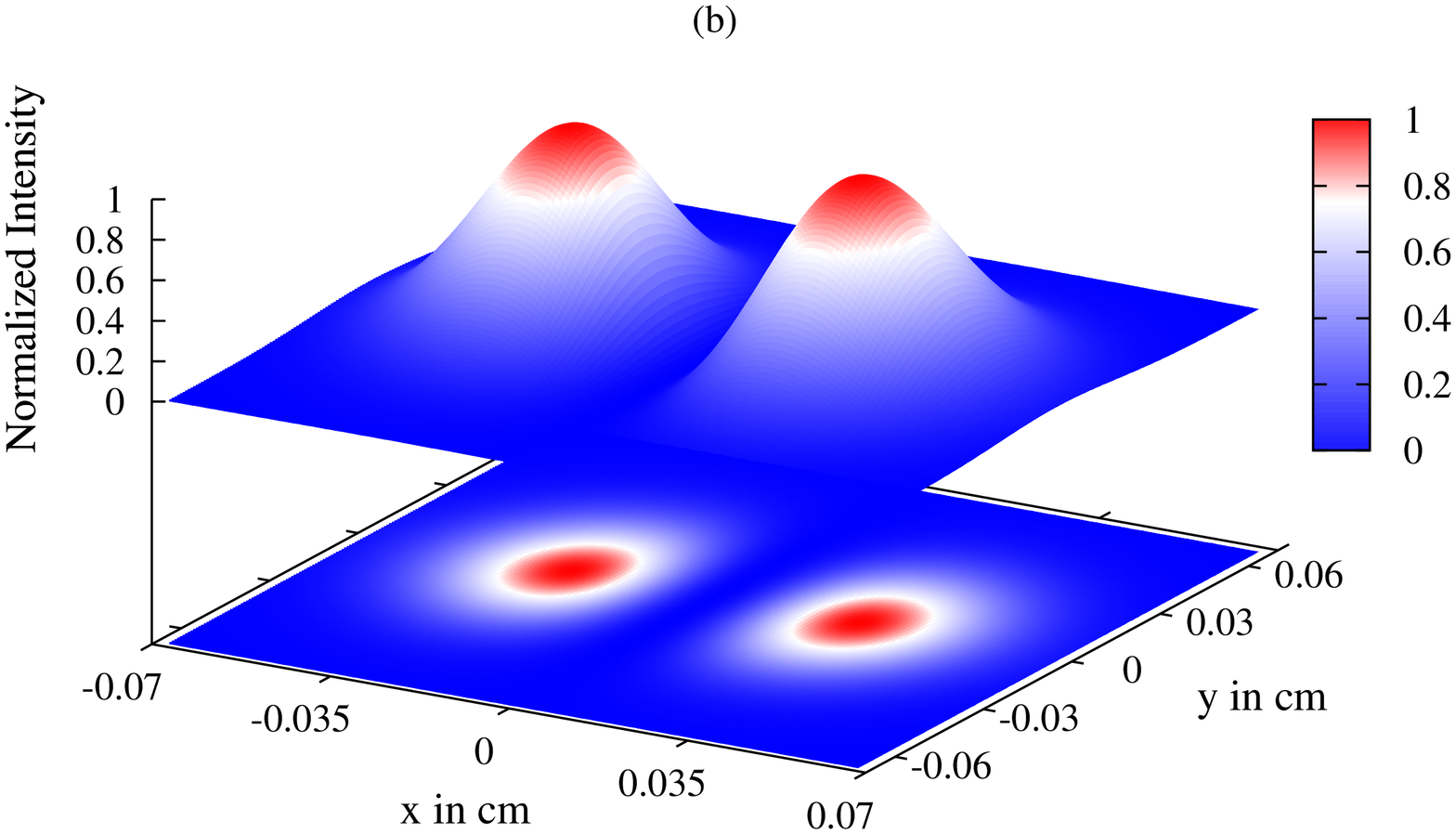}
\caption{\label{fig:Fig7}(Color online) Transverse spatial intensity profile of the probe and control fields after propagation through an atomic vapor cell of length 4cm. 
The initial control field profile is chosen as a Hermite-Gaussian doubly-peaked profile. The top panel (a) shows the probe field, the bottom panel (b) for the control field. 
The parameters are as in Fig.~\ref{fig:Fig4}. }
\end{figure}
\begin{figure}[t]
\includegraphics[width=1.0\columnwidth]{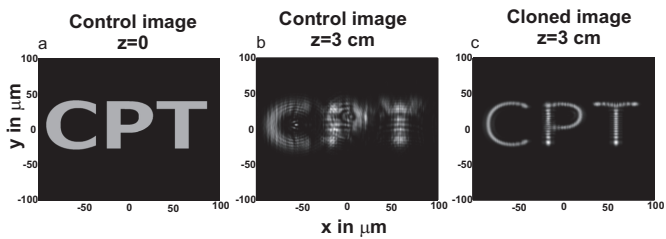}
\caption{\label{fig:Fig9} Transverse spatial intensity profile of the control and probe fields. (a) shows the initial profile of the control field at the medium entry. (b) shows the control field profile after propagation through a vapor cell of length 3cm filled with rubidium atoms. (c) shows the probe field profile after the propagation. Initially, the probe field is chosen as a plane wave.
The parameters are $g=0.15{\gamma}$, $G=1.5{\gamma}$,  $\DC=0$, $\DP=0.002{\gamma}$, $\Gamma=0.001\gamma$, and $N=10^{12}$ atoms/cm$^{3}$.}
\end{figure}

Next, we study cloning of a two-peaked Hermite-Gaussian control beam onto the spatial profile of the probe beam. Fig.~\ref{fig:Fig7} shows the peak-normalized intensity of both the transmitted probe and control beams after a propagation distance of $4$~cm. It can be seen from Fig.~\ref{fig:Fig7} that while the overall structure of the control beam is preserved, the width of the cloned probe beam profile is reduced by a factor of about $2.5$ as compared to the initial width of the control beam. Therefore, the finesse of the transmitted probe beam is about 2.5 times greater than that of the control beam, which is consistent with the findings of an recent experiment by Li {\it et al} \cite{Li}. The integrated transmission intensity of the cloned output probe beam is $5\%$ of the integrated intensity of the input probe beam. The transmission of the cloned beam can be increased by decreasing the optical density. We also found that the width of the cloned beam can be decreased by decreasing the relative intensity 
of the probe and control field giving us control over the finesse of the cloned probe beam. 

Finally, Fig.~\ref{fig:Fig9} shows how a complex image encoded in the transverse control field intensity profile can efficiently be cloned onto the probe field. For this, we choose the three letters ``CPT'' as initial profile of the control beam, and a plane wave for the initial probe beam profile.  We find from Fig.~\ref{fig:Fig9}(b) that the control beam profile is significantly distorted after 3 cm propagation length through  a vapor cell. This distortion arises from diffraction and from the variation of the refractive index experienced by the control field. 
In contrast, the fiber-like spatial dispersion of the probe field can be used to compensate or to reverse the effect of the diffraction. As a result,  the cloned image in the transverse profile of the probe beam is tightly focused. We further note from  panel (c) that the feature size in the cloned image is reduced by a factor of 2 as compared to the incident control field image. This suggests that the finesse of the transmitted cloned image is two times greater then that of the incident control image. Hence, the feature size of the cloned image is decreased twofold. 

\section{\label{Conclusion}Conclusion}

In conclusion, we have studied the possibility of cloning of an arbitrary images encoded in the spatial profile of a control beam onto a probe beam. Our method is based on coherent population trapping in a three-level atomic lambda system. We have considered both, weak and strong probe fields with strength comparable to the control field, and have discussed the differences for light propagation in the weak and strong field limits.
In the ``strong field limit'' where the probe and control fields are of comparable strength, we firstly calculated the atomic susceptibilities including both linear and nonlinear effects for the two fields. We then found that a waveguide-like structure can be formed inside the medium at red detuning of the probe field when applying a Gaussian control and a super-Gaussian probe. 
At the same time, a transparency window centered in the waveguide, which can be controlled by changing the relative intensities of the probe and control fields. This transparency window allows to transfer the transverse intensity profile of the control onto the transmitted probe field. By numerically solving the propagation equations for both fields, we found that the spatial profile of the control is gradually mapped onto the transmitted probe already after a few Rayleigh lengths. Interestingly, the feature size in the spatial profile of the probe field is reduced compared to that of the original control field structure.  In order to show that our method works for  arbitrary images, different spatial profiles of the control and probe fields are considered. In particular, the three letters ``CPT" initially encoded on the control field are cloned onto the transmission profile of the probe. In this process, the feature size is decreased by a factor of 2. Interestingly, even though cloning of arbitrary images 
onto the probe field is constructed nicely,  the control field image is severely distorted due to diffraction throughout the propagation. 

\begin{acknowledgements}
{T.N.D gratefully acknowledge funding by the Science and Engineering Board(SR/S2/LOP-0033/2010). L. Z. and J. E. are grateful for funding by the German Science Foundation (DFG, Sachbeihilfe EV 157/2-1).}

\end{acknowledgements}

\appendix
\section{\label{app-A}Coefficients for susceptibility}

\begin{widetext}
\begin{align}
N_{31}=&(|G|^2({\gamma}(i{\gamma}+{\DP})({\Gamma}^2+({\DC}-{\DP})^2)
 +({\gamma}(i{\Gamma}+{\DC}-{\DP})+{\Gamma}({\DC}+{\DP}))|g|^2+{\gamma}(i{\Gamma}+{\DC}-{\DP})|G|^2)g)\\
N_{32}=&(|g|^2({\gamma}(i{\gamma}+{\DC})({\Gamma}^2+({\DC}-{\DP})^2)+{\gamma}(i{\Gamma}-{\DC}+{\DP})|g|^2
 +({\gamma}(i{\Gamma}-{\DC}{\DP})+{\Gamma}({\DC}+{\DP}))|G|^2)G)\\
D=&\gamma |G|^6 +|G|^4\left[3 |g|^2(\gamma +2 \Gamma )+2 \gamma(\gamma  \Gamma +\DP(\DC-\DP))\right]\notag\\
&+\gamma  | g|^2 \left[2 |g|^2(\gamma \Gamma +\DC(\DP-\DC))+|g|^4+\left(\gamma^2+\DC\right)((\DC-\DP)^2+\Gamma ^2)\right]\notag\\
&+|G|^2 [| g|^2 \left((4 \gamma +\Gamma ) \DC^2+2 \gamma  \Gamma  (2 \gamma +3 \Gamma )+2 (\Gamma -4\gamma ) \DC \DP+(4\gamma +\Gamma )\DP^2\right)\notag\\
&+3| g|^4 (\gamma +2 \Gamma )+\gamma(\gamma ^2+\DP^2)((\DC-\DP)^2+\Gamma ^2)]\,.
\end{align}
\end{widetext} 

\end{document}